\documentclass[prl,twocolumn,superscriptaddress]{revtex4}

\begin{document}


\title[]{Comment on ``Unique Translation between Hamiltonian Operators and Functional Integrals"}


\author{Michael Weyrauch}
\email[]{michael.weyrauch@ptb.de}
\affiliation{Physikalisch-Technische Bundesanstalt, D-38116
Braunschweig, Germany}
\author{Andreas W. Schreiber}
\email[]{aschreib@physics.adelaide.edu.au} \affiliation{Center
for the Subatomic Structure of Matter (CSSM), Adelaide, SA 5001,
Australia}



\date{\today}

\begin{abstract}
\end{abstract}

\maketitle


In a recent letter\cite{gollwett}, Gollisch and Wetterich (GW)
show that a careful treatment of discretization errors in
a phase-space path integral formulation of quantum mechanics leads to a
correction term as compared to the standard form based on
coherent states. Since the coherent state formalism is widely used in field theory 
and statistical physics, one would have to view any suggestion that it may produce incorrect results with deep concern. In order to support their finding, GW study the
simple one-dimensional nonlinear oscillator described by the
Hamiltonian
\begin{equation}\label{hamiltonian}
H= m a^\dagger a + \frac{\lambda}{2}a^\dagger a^\dagger a a
\end{equation}
and calculate the thermal expectation value $\langle p^2\rangle$
both in conventional quantum mechanics and the proposed new
functional integral approach ($p=i(a^\dagger-a)/\sqrt{2}$). A
comparison of both calculations {\it indeed} supports the proposed correction
term. However, in this Comment, we point out that a calculation of $\langle p^2\rangle$
within the standard coherent state path integral approach
{\it also} agrees with conventional quantum
mechanics, provided that discretization errors are correctly controlled here as well.

Standard coherent state path integral techniques~\cite{Negele}
allow us to write down the partition function $Z$ for the
Hamiltonian~(\ref{hamiltonian})
\begin{equation}
Z=\lim_{N \rightarrow\infty} \int_{z_0=z_N}   \left( \prod_{k=1}^N
\frac{d^2z_k}{\pi} \right) \exp(-S)
\end{equation}
with the discrete representation for the action
\begin{equation}
S=\sum_{k=1}^N z_k^*(z_k-z_{k-1})+\frac{\beta
m}{N}z_k^*z_{k-1}+\frac{\beta\lambda}{2N} z_k^{* 2}z_{k-1}^2.
\end{equation}
Here, $\beta$ denotes inverse temperature.
We now transform into Fourier space using
\begin{equation}
z_k= \sum_{n=-N/2+1}^{N/2} \phi_n \exp(2\pi i n k/N).
\end{equation}
In the new integration variables $\phi_n$, the action reads
\begin{eqnarray}\label{wirk}
S&=&\sum_{n=-N/2+1}^{N/2}\phi_n^*\phi_n N[1-e^{-\frac{2\pi i
n}{N}}(1-\frac{\beta m}{N})] \nonumber\\
&+&\frac{\beta\lambda}{2}\sum_{n,n_1,n_2} \phi_{n_1-n}^*
\phi_{n_2+n}^* \phi_{n_1} \phi_{n_2}e^{-\frac{2\pi i}{N}(n_1+n_2)}
\end{eqnarray}
with the continuum limit ($N\rightarrow\infty$, $\epsilon = 1/N$)
\begin{eqnarray}\label{cont limit}
S &=&\sum_{n=-\infty}^\infty\phi^*_n\phi_{n}(2\pi i n + \beta m
) \nonumber\\
&+&\frac{\beta\lambda}{2} \sum_{n,n_1,n_2} \phi_{n_1-n}^*
\phi_{n_2+n}^* \phi_{n_1} \phi_{n_2}e^{-2\pi i (n_1+n_2)
\epsilon}.
\end{eqnarray}
It is understood here that the integration measure of the functional integral is suitably normalized so that the correct free particle limit of the theory is obtained.
Eq.~(\ref{cont limit}) does not contain the correction term
proposed by GW in their eq.~(11). Instead, it contains an
exponential factor, which guarantees convergence at large n. 
The limit $\epsilon \rightarrow 0^+$ must be taken only after the evaluation of the functional integral.  Although sometimes
neglected (as e.g. in Ref.~\cite{schulmann}),
the importance of this exponential factor is 
pointed out, for example, in Refs.~\cite{Negele,Klauder,LuttWard}.

With the action~(\ref{cont limit}), a standard calculation in first
order perturbation theory yields
\begin{equation}\label{zet}
Z=\frac{1}{1-e^{-\beta
m}}\left[1-\beta\lambda\left(\frac{e^{-\beta m} }{1-e^{-\beta
m}}\right)^2\right]
\end{equation}
from which one easily obtains
\begin{equation}\label{pq}
\langle p^2 \rangle = \frac{1}{2}\coth \frac{\beta
m}{2}-\frac{\beta\lambda}{4}\frac{e^{-\beta m/2}}{\sinh^3 (\beta
m/2)}.
\end{equation}
A simple analysis confirms that eqs.~(\ref{zet}) and~(\ref{pq})
exactly agree with conventional quantum mechanics as well as the
GW results up to order $\lambda$ (cf. eq.~(14) in Ref.~\cite{gollwett}). 

To conclude, we agree that the phase-space path integral approach proposed by GW,
which employs a particular operator ordering
resulting in a correction term to the continuum action, provides {\em a correct} translation between 
Hamiltonian operators and functional integrals.  However, as pointed out in this Comment, this approach
{\em is not} unique. The coherent state path integral formalism, without correction term,  yields the same
results due to exponential convergence factors to which we have drawn attention above. 

One of the authors (MW) would like to thank the CSSM for its
hospitality and Walter Apel, Janusz Szwabi\'nski, and Wolfgang
Weller for helpful discussions.

\end{document}